\documentclass[a4paper,11pt]{article}
\pdfoutput=1 

\usepackage{jcappub} 

\usepackage[T1]{fontenc} 
\usepackage{color}

\title{\boldmath Possible role of magnetic reconnection in the electromagnetic counterpart of binary black hole merger}

\author{F. Fraschetti} 

\affiliation{Departments of Planetary Sciences and Astronomy, University of Arizona, Tucson, AZ, 85721, USA} \affiliation{Harvard/Smithsonian Center for Astrophysics, 60 Garden Street, MS-06, Cambridge MA 02138, USA}

\emailAdd{ffrasche@lpl.arizona.edu}

\abstract{We propose a qualitative scenario to interpret the argued association between the direct measurement of the gravitational wave event GW150914 by Laser Interferometer Gravitational Wave Observatory (LIGO)-Virgo collaborations and the hard $X$-ray transient detected by Fermi-Gamma-ray Burst Monitor (GBM) $0.4$ sec after. In a binary system of two gravitationally collapsing objects with a non-vanishing electric charge, the compenetration of the two magnetospheres occurring during the coalescence, through magnetic reconnection, produces a highly collimated relativistic outflow that becomes optically thin and shines in the GBM field of view. We propose that this process should be expected as a commonplace in the future joint gravitational/electromagnetic detections and, in case of neutron star-neutron star merger event, might lead to detectable $X$- or $\gamma$-ray precursors to, or transients associated with, the gravitational bursts.
}

\begin{document}
\maketitle
\flushbottom

\section{Introduction}
\label{sec:intro}

The long-awaited direct detection of a gravitational waves source (GW150914) by the LIGO collaboration with the Virgo collaboration \citep{Abbott.etal_PRL:16} was shown to be consistent with numerical simulations of the merger of a black hole (hereafter BH) binary system with masses $\sim 30 M_\odot$ \citep{Abbott.etal_ApJ:16}. It has been already stressed that this result is groundbreaking in several respects: it shows that binaries of gravitationally collapsed objects exist and merge within the age of the universe. The merger of binary supermassive BHs (typically $M \sim 10^8 M_\odot$) has been shown by a number of numerical simulations \citep{Palenzuela.etal:10,Alic.etal:12} to produce a collimated electromagnetic flash accompanying the GW event. However, scalings from the pioneeristic work by \cite{Blandford.Znajek:77} are inconsistent with the large luminosity ($10^{49}$ erg$/$s) of the transient signal detected by GBM on board the {\it Fermi} satellite at photon energy $>50$ keV \citep{Connaughton.etal:16}, as discussed by \cite{Lyutikov:16}.
Owing to the exceptional impact of such an association, it seems worth exploring at least qualitatively a possible scenario for the joint events as already done, e.g., in \cite{Loeb:16,Zhang:16,Perna.etal:16}.

The widely accepted Blanford-Znajek mechanism envisages supermassive rotating BHs powering electromagnetic jets with luminosity $L \sim 10^{45}$ erg$/$s and with circum-BH disks currents sustaining a steady magnetic field near the BH of order of $10^4$ gauss. The luminosity of the GBM transient combined with $30 M_\odot$ BHs makes the association of GW150914 with the GBM transient unlikely as it would imply implausibly high magnetic fields. We propose that the compenetration of the two magnetospheres amplifies by orders of magnitude the magnetic energy and drive, through magnetic reconnection, the release of a relativistically hot and optically thick pair plasma. We argue in this short note that the latter scenario might be a commonplace electromagnetic source associated with merger of collapsed or compact objects, such as BHs or neutron stars.

\section{Model}\label{Non-diff}

\begin{figure}
\includegraphics[width=13cm]{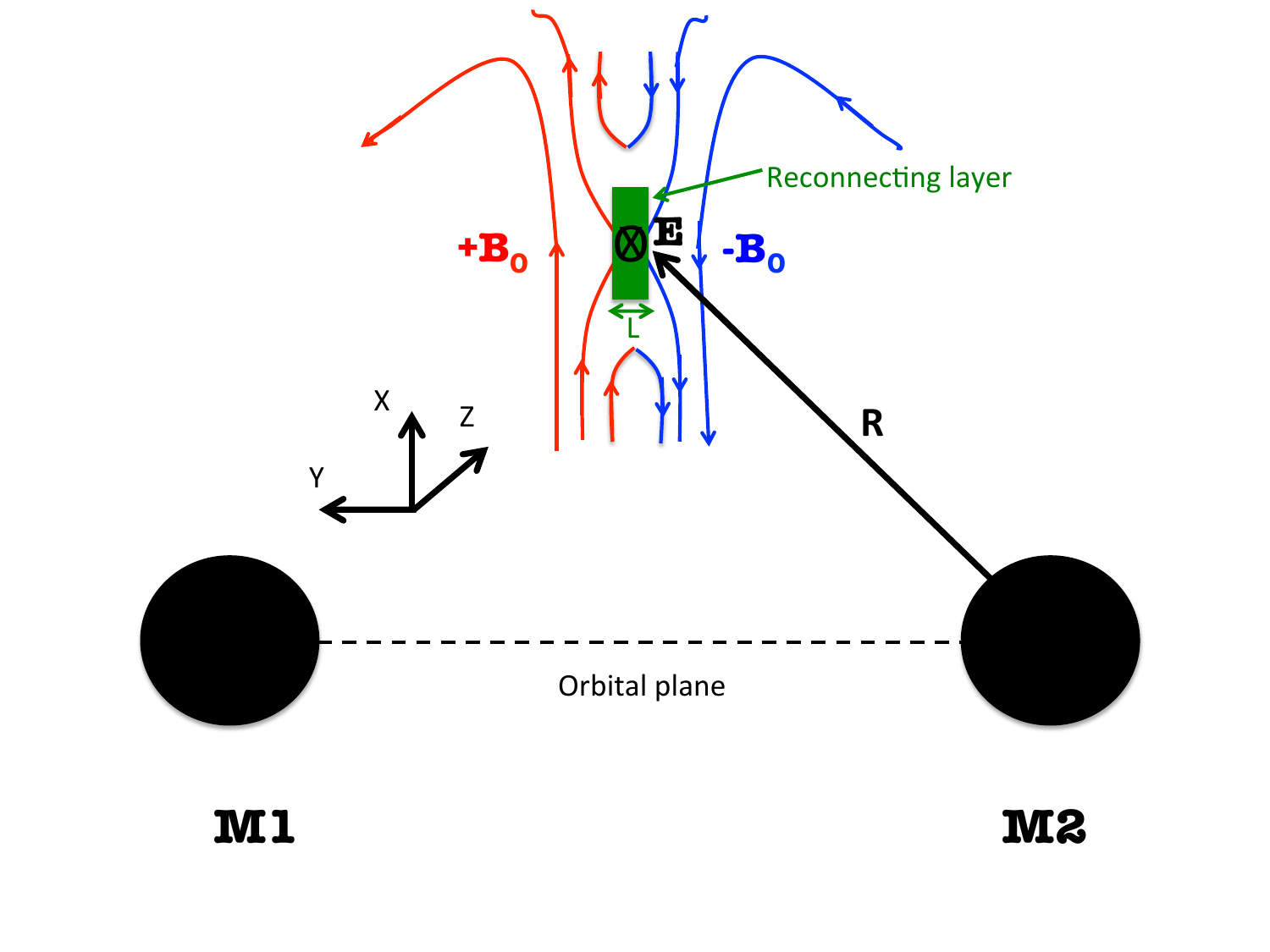}
\caption{Cartoon illustration of the current sheet region ejecting relativistically hot flow along $x$-axis and charged particles that move back and forth across the reconnection layer while being accelerated by $E_z$ along the Speiser orbits. \label{fig}}
\end{figure}

The source of GW 150914 very likely comprises of two objects at the final stage of their gravitational collapse, close to their respective horizons, inspiralling and moving at a good fraction of the speed of light. Even if mass ($M$), charge ($Q$) and spin angular momentum ($S$) are BHs observables, it is customary to assume that stationary BHs in the universe have no substantial charge ($Q/\sqrt{G}M \ll 1$) for a far-away observer as it will be quenched on a presumably short time scale whereas $M$ and $S$ evolve on a much longer time scale. We allow for each of the collapsing objects to carry a non-vanishing global electric charge $Q/\sqrt{G}M \ll 1$ and $S$. If the two spins $S$ are anti-aligned and approximately orthogonal to the orbital plane, the compression of the anti-parallel magnetic field lines within the current sheet separating the two magnetospheres will drive during the inspiral a violent magnetic reconnection, which can eject a relativistically hot pair plasma that later emits the hard $X$-rays signal detected by Fermi (see the cartoon illustration in Fig. \ref{fig}).     

The problem of gravitational collapse of an electrically charged and spherically symmetric shell has been long investigated \citep{Israel.66}. Astrophysical implications of a charged collapsing object have been explored in the past two decades in connection with Gamma-Ray Bursts (GRBs) prompt and afterglow high-energy light curves (see, e.g., \cite{Ruffini.etal:01b,Ruffini.etal:05}) and $X-$ and $\gamma$-ray spectral evolution \cite{Fraschetti.etal:06}. 

For a stationary isolated Kerr-Newman (hereafter KN)\footnote{Kerr-Newmann space-time asymptotic solution can be used here only as a reference since it does not apply to this case because the two merging systems are not isolated and the surrounding space-time of each system is largely deformed by the companion.} BH, it is well-known that, upon developing the general relativistic solution in a far-away observer rest frame, the leading components to order $1/r$ of the electromagnetic stress-energy tensor $F_{\mu \nu}$ behave, as for the magnetic field $B$, as magnetic dipole \cite{Misner.Thorne.Wheeler:73}: 
\begin{equation}
B'_r = F_{\theta \phi} \sim { 2Q'a' \over r^3} {\rm cos }\theta ,  \quad  B'_\theta = F_{\phi r} \sim { Q'a' \over r^3} {\rm sin }\theta, \quad B'_\phi = F_{r \theta} \sim 0 ,
\label{B}
\end{equation}
where $a'=S/M $ is the angular momentum of spin per mass unit, all primed quantities are in geometric units where $G=c=1$ and $t,r,\theta,\phi$ are Boyer-Lindquist \cite{Boyer.Lindquist:67} coordinates\footnote{For an observer closer to the BH, the locally non-rotating coordinates frame, introduced by \cite{Bardeen.etal:72} for the Kerr metric (so-called ``Zero Angular Momentum Observer'' or ZAMO), could be used. The ZAMO frame has the advantage of co-rotating with the BH as a result of frame dragging only. In this work expressing $B$ in the ZAMO would lead to unnecessary complicated dependences on $r$ and $(Q/M)^2$; due to the rapid drop of $B$ with $r$ and to the small inferred charge ($(Q/M)^2 \ll 1$ as shown in Sect. \ref{Discussion}), ZAMO is not used in Eq. \ref{B}.}. The space-time geometry within a binary BH circular orbit is accessible only through numerical simulations and an analytical formulation is not known to date. In addition, the LIGO/Virgo signal is consistent with fastly rotating BHs, therefore perturbative techniques do not apply. Our first ansatz is that the magnetic dipole structure is approximately preserved within the inspiralling orbit at a certain latitude from the orbital plane where reconnection occurs. In a forthcoming work we will describe the kinetic process within the reconnection layer in a suitable local frame.

In Eq. \ref{B} the charge $Q'$ can be expressed in terms of the extreme charge, ${Q'_{ex}}$, for a KN-BH. The value of ${Q'_{ex}}$ is determined from the existence of the event horizon $r_+$ ($r
_\pm = M' \pm \sqrt{M'^2 -{Q'}^2 -a'^2}$), namely ${Q'_{ex}}^2 =M'^2 - a'^2$. By using the dimensionless spin angular momentum $a = S c /G M^2$, 
\begin{equation}
{Q_{ex}^2} = G M^2 (1 - a^2)
\label{Qex2}
\end{equation}
where unprimed quantities are in CGS units. Equation \ref{Qex2} leads to 
\begin{equation}
{Q_{ex}} = 5.11 \times 10^{30} \left( \frac{M}{10 M_\odot} \right) \sqrt{1 - a^2} \, {\rm esu} =  1.7 \times 10^{21} \left( \frac{M}{10 M_\odot} \right) \sqrt{1 - a^2} \, {\rm Coulomb}
\label{Qex4}
\end{equation}
The luminosity of the GBM transient constrains the value of $Q$ (see Sect. \ref{Discussion}).

Magnetic reconnection has been repeatedly invoked in the past decade to explain the production of non-thermal particles populations, beside thermal heating, for those observations inconsistent with constraints from other mechanisms or with ideal MHD, traditionally in solar flares \citep{Parker:57} and more recently in the intensity peak of energetic particles at solar termination shock \citep{Zank.etal:15}, and in relativistic astrophysical outflows (see \cite{Kagan.etal:15} and references therein). 

We argue that, during the last phase of the magnetospheres coalescence, the compenetration at the current sheet layer dissipates magnetic energy into relativistically hot thermal electron-positron pair plasma, possibly baryon-loaded (and a certain amount of non-thermal particles, not relevant to the discussion here) ejected in a highly collimated relativistic outflow. 
Figure \ref{fig} shows the current sheet layer that ejects the hot thermal pair plasma into the downstream cone along $x$-axis. The electric field $E_z$ accelerates particles (electrons along $-z$ and positrons along $+z$) as long as the particles remain confined within the boundary layer and might give rise to a non-thermal distribution (see, e.g., recent simulations of relativistic reconnection in flat space-time in \cite{Cerutti.etal:13}). Assuming $B_z =0$ does not alter the main process in this simplified scenario.
 
The relativistic highly collimated thermal pair plasma, and photons produced by their annihilation, outflowing the reconnecting region will propagate (bulk Lorentz $\Gamma \sim 100$), possibly preceded by a shock, as far as the plasma rest frame opacity of the high-energy $\gamma$-ray to the electron-positron pair production exceeds unity (similar scenario is currently the most credited description of GRBs \cite{Piran.Shemi:93}). At such point the electromagnetic radiation is released in a single pulse detected at Earth as the GBM transient of arrival time $t_a$ duration given by $\Delta t_a \sim 1$ sec. The GW signal emitted at the binary merger travels undisturbed at the speed of light \cite{Abbott.etal_LIGO_Virgo_Fermi_Integral:17} with infinite mean free path. It has been already estimated \citep{Zhang:16} that, if GRBs model are to be used as a guide \citep[e.g.,][]{Ruffini.etal:01b,Kumar.Zhang:15}, the observed arrival time delay $\Delta t_a \simeq 0.4$ sec of the GRB signal with respect to the GW signal is compatible with a distance travelled by an ultra-relativistic optically thick plasma ($\Gamma \sim 100$) of order of $10^{14}$ cm. We note that $\Delta t_a$ can be stretched if the plasma is baryon-loaded. At such a distance from the coalescence location, much larger than the clean circum-merger medium, the ejected plasma emits the GBM flash\footnote{If the relativistic flow ejected from the reconnecting layer is baryon-loaded, an afterglow at longer wavelengths might be expected.}.

\section{Discussion}\label{Discussion}

The scenario qualitatively described above requires obviously a detailed demonstration through analytic calculations or extended resistive general relativistic MHD simulations. We only argue here that the process of magnetic reconnection at the compenetration of magnetospheres of compact objects might be considerable source of electromagnetic radiation. The reconnection growth rate in relativistic force-free approximation \citep{Lyutikov:03} can be cast as 
\begin{equation}
\gamma \simeq 1/\sqrt{\tau_R \tau_A} ,
\end{equation}
where $\tau_A$ is the Alfv\'en time-scale and $\tau_R$ the resistive time-scale, both measured in the local  frame. For a current sheet thickness $L = L_{6} 10^6$ cm, we can write $\tau_R \simeq L^2/\eta$ and $\tau_A \simeq L/c$, where the plasma resistivity $\eta$ can be expressed in terms of the electron skin depth $\delta = c/\omega_{p,e}$ as $\eta \simeq c \, \delta$ (here $\omega_{p,e}= \sqrt{4 \pi n_e e^2/m_e}$ is the electron plasma frequency and $e$, $n_e$ and $m_e$ charge, number density and mass of electrons, respectively). For $n_e \simeq 10^{10}$ cm$^{-3}$, much smaller than numerically determined densities in proximity of jet-powering BHs \cite[e.g.][]{Barkov.Komissarov:08}, one can approximate the reconnection time scale as \citep[cfr.][]{Lyutikov:03}
\begin{equation}
\tau = {1\over \gamma} \sim \left({L \over c}\right)^{3/2} \sqrt{\omega_{p,e}} \sim 10^{-2} L_{6}^{3/2} \, {\rm sec} .
\label{tau_eq}
\end{equation}
The time-scale $\tau$ is measured in the local reference frame of the reconnecting layer; thus, it does not account for the gravitational redshift in the unknown space-time geometry. We remind that the frequency at the observer frame $\nu_{obs} $ is shifted from the frequency in the source frame $\nu_s$ by the lapse factor $\alpha$ that in ZAMO-frame reads $\alpha = \nu_{obs} /\nu_s =\Delta^{1/2} \sin \theta /\sqrt{g_{\phi \phi}}$, where $\Delta =(r - r_+)(r-r_-)$ and $g_{\phi \phi}$ is the $(\phi \phi)$ covariant component of the KN metric tensor\footnote{In geometric units $g_{\phi \phi} = [r^2 + a^2 +(2Mr-Q'^2)a^2 \sin^2 \theta / \rho^2]\sin^2 \theta$ where $\rho^2 = r^2 + a^2 \cos \theta$.} ($\alpha \rightarrow 0,1$ for $r \rightarrow r_+ , \infty$). 
However, on the very short merging time-scale due to highly relativistic flow speed involved, the actual reconnection can be limited to a current sheet thickness $L \ll 10^6$ cm, reducing $\tau$ to sub-millisecond scale \citep{Lyubarsky:05}. The conversion of $\tau$ to the observer frame is also affected by Doppler shift (${\cal D} = \Gamma [1-(v/c) \cos \vartheta]$, where $v$ is the source speed and $\vartheta$ is the angle of the direction of photon emission with respect to the line of sight to the observer located far away from the source), arrival time dilation due to the fact that two photons emitted with a certain time-lag $\Delta t$ in the plasma comoving frame are detected with a different time-lag by the observer at Earth $\Delta t_a$ because the source is moving ($\Delta t_a \sim \Delta t /2 \Gamma^2$) and cosmological redshift ($z \sim 0.09$ for GW150914). 
Such estimates show that a delayed pulse at $0.4$ sec from the GW150914 can easily be produced.

We assume that the magnetic field $B_0$ upstream of the reconnecting current sheet inflows at speed $\sim c$ in the late merging phase at some distance $R$ from each mass. The inflowing magnetic energy per unit time is efficiently converted into the electromagnetic luminosity of the GBM transient ($L_{EM} \simeq 10^{49}$ erg$/$s) if
\begin{equation}
{B_0^2 \over 8\pi} c R^2 \simeq L_{EM} .
\label{B_L}
\end{equation}
A reconnecting layer located at 
\begin{equation}
R = R_0 {GM \over c^2} = 1.5\times 10^7 {\rm cm} \, \left({R_0 \over 10}\right) \left({M \over {10 \, M_\odot}}\right) 
\end{equation}
leads to a magnetic field $B_0 \simeq 2 \times 10^{12}$ gauss ({$R \simeq 6  r_+$ so that} $R^3 \gg  r_+^3$, validating the expression of $B$ in Eq. \ref{B}). 
For a reconnection layer at smaller $R \gtrsim r_+$, $B$ has to be calculated in, e.g., ZAMO frame, unlike Eq. \ref{B}.

The net charge $Q$ carried by each of the collapsing objects can be determined as follows. By assuming that the magnetic field of an isolated KN-BH in Eq. \ref{B} is a good approximation of the total field at distance $R$ where reconnection occurs, the magnetic energy density is given by
\begin{equation}
{B_0^2 \over 8 \pi}  
\simeq {1 \over 4 \pi} \left({Q a' \over r^3} \right)^2 \, ,
\label{B_Q}
\end{equation}
where we have used $B_0 = B'_0 c^2/\sqrt{G}$ and $Q = Q' c^2/\sqrt{G}$ and we have approximated $\cos^2  \theta \sim 1/3$ since the reconnection reasonably occurs at mid-latitude during the inspiral.
By eliminating $B_0$ from Eq.s \ref{B_L} and \ref{B_Q} and calculating Eq. \ref{B_Q} at $r=R$, we can express  $Q$ in terms of $Q_{ex}$ in Eq. \ref{Qex2}: 
\begin{equation}
\left (Q \over Q_{ex} \right)^2 = 
4 \pi {G \over c^5} L_{EM} \frac{R_0^4}{a^2 (1-a^2)} = 3.45 \times 10^{-10} \, L_{EM}^{49} \frac{R_0^4}{a^2 (1-a^2)} \, ,
\end{equation}
where $L_{EM} = L_{EM}^{49} \times 10^{49}$ erg/s. Thus, the charge $Q$ can be recast as
\begin{equation}
Q = 3.15 \times 10^{16}  \sqrt{L_{EM}^{49}}  \frac{R_0^2}{a}  \frac{M}{10 M_\odot} \, {\rm Coulomb} .
\label{Q_fin}
\end{equation}
For the parameters inferred by the joint LIGO/Virgo measurement and simulations \cite{Abbott.etal_PRL:16}, i.e. $M = 30 M_\odot$, $a \simeq 0.7$ (although such a value is an upper limit), and for $L_{EM}^{49} = 1.$, we find
\begin{equation}
Q = 3.7 \times 10^{-3} Q_{ex} = 1.3 \times 10^{19} \, {\rm Coulomb}
\end{equation}
Equivalently, by using Eq. \ref{Qex2}, we conclude that the $B_0$ needed to explain the $X$-ray transient would lead to a change in the KN metric tensor only of order $Q^2/G M^2 \simeq 7 \times 10^{-6}$, much smaller than $a^2 \simeq 0.5$.

Owing to the very large proton charge-to-mass ratio ($\sim 10^{18}$), charged BHs face instability due to charge-neutralization. However, the net charge $Q$ needs to sustain the merging magnetospheres in typical conditions for the interstellar medium only during the reconnection event. In other words, the time interval needed for such $Q$ to exist in the local frame at the reconnecting layer, $\delta \tau_{loc}$, has to exceed $ \tau$. If we use, for instance, the lapse $\alpha$ between a ZAMO $\delta \tau_{loc}$ and a far-away observer $\delta \tau_{obs}$ ($\alpha = \delta \tau_{loc}/\delta \tau_{obs}$), we find that $\alpha \gtrsim 0.8$ at $r \gtrsim R$. Thus, the gravitational redshift at $r \gtrsim R$ introduces a little correction as compared with the uncertainty on $L$; as a result, a net charge is needed for a $\delta \tau_{obs}$ in the milli-second range for ($L_6 \sim 0.1$). We note that the time scale $R/c \simeq R_0 GM/c^3 \sim$ ms, larger by a factor $R_0 =10$ than the time-evolution of gravitational collapse ($GM/c^3$), is consistent with numerical simulations of the evolution of a few tens $M_\odot$ rotating collapsing stars \cite{Lehner.etal:12}. The lower limit of the discharge time scale therefore is $R/c \simeq R_0 GM/c^3$. Finally, such a short $\delta \tau_{obs}$ suggests that several events of reconnection might take place during a single merger event, each of them powering an observable outflow in a certain direction. 

We do not discuss here the process leading to the formation of each collapsing rotating star. Before the detection of GW150914, binary BHs were not deemed to exist in reality: models of gravitational collapse of massive stars in an already formed binary system and model of coalescence of two BHs moving one with respect to the other with a sufficiently small impact parameter have both unresolved issues. We have used here magnetosphere of a KN-BH, expected to represent theoretically the physical asymptotic state of such process of gravitational collapse. 

It is expected that in the ZAMO frame where the reconnecting magnetic field lines come close to each other, the current sheet layer is dragged along the orbital motion of the compact objects and stretched by differential rotation, developing strong shear instabilities (see \cite{Corvino.etal:10} for the shear instability for an isolated neutron star). Those instabilities will presumably grow faster and faster as the system merge.

This scenario will be further investigated within either the framework of tearing mode instabilities, or Petschek model of two slow mode shocks \cite{Lyubarsky:05}, both cases in the incompressible regime as usually assumed in the literature.

\section{Conclusion}\label{Conclusion}

We have proposed a qualitative scenario to account for the putative association between the first direct detection of a gravitational wave signal (GW150914), interpreted as a merger of two fast-spinning $30 M_\odot$ black-holes, and a hard $X-$ray transient detected $0.4$ sec after the GW event. If the objects can carry electric charge of order $0.1 \%$ of the extremal charge for the spin angular momentum inferred from simulations, the compenetration of the two magnetospheres can lead to an optically thick relativistic outflow via magnetic reconnection. The transparency of such plasma will shine in the $X$- and $\gamma$-ray detectors within $\sim 1$ sec from the GW signal. We emphasize that several episodes of reconnection are expected to occur during the BH-BH mergers leading to several $X$-ray transients, possibly precursors, for the gravitational wave signal at detectors.

The recent detection on August 17 2017 of the gravitational wave event GW170817 by Advanced LIGO and Virgo detectors and, independently, of the short GRB 170817A by Fermi/GBM and Integral opens new prospect of detectability of the reconnection event proposed here. During the GW signal in the sensitivity band ($\sim 100$ sec for GW170817), a number of high-energy precursors could be detected, although constraining upper limits are set in the case of GW170817 \cite{Goldstein.etal:17b}. We hope that the scenario here proposed will stimulate fresh thoughts and prompt detailed resistive general relativity numerical investigations.  


\acknowledgments

A constructive feedback from the referee is acknowledged. We acknowledge useful comments and feedbacks by D. Arnett, M. Begelman, M. Rees, L. Rezzolla, E. Vishniac, B. Zhang. This work was supported, in part, by NASA under Grants  NNX13AG10G and NNX15AJ71G.  


\def \apss{{\it Astrophys.\ Sp.\ Sci.}}
\def \aj{{\it The Astronomical Journal}}
\def \apj{{\it The Astrophysical Journal}}
\def \apjl{{\it The Astrophysical Journal Letters}}
\def \apjs{{\it The Astrophysical Journal Supplement Series}}
\def \araa{{\it Annual Review Astronomy \& Astrophysics}}
\def \prc{{\it Physical\ Review\ C}}
\def \prd{{\it Physical\ Review\ D}}
\def \aap{{\it Astronomy \& Astrophysics}}
\def \aaps{{\it Astronomy \& Astrophysics Supplement Series}}
\def \gca{{\it Geochim. Cosmochim.\ Act.}}
\def \grl{{\it Geophysical Research Letters}}
\def \jgr{{\it Journal of Geophysical Research}}
\def \mnras{{\it Monthly Notices of the Royal Astronomical Society}}
\def \nat{{\it Nature}}
\def \physscr{{\it Physica\ Scripta}}
\def \pre{{\it Physical\ Review\ E}}
\def \physrep{{\it Physics Report}}
\def \planss{{\it Planetary and Space Science}}
\def \pasp{{\it Publ.\ Astron.\ Soc.\ Pac.}}
\def \solphys{{\it Solar\ Physics}}
\def \ssr{{\it Space\ Science\ Reviews}}



\bibliographystyle{JHEP}
\bibliography{ffraschetti}



\end{document}